\documentclass[11pt,final]{article}
\usepackage{fancyvrb, color, graphicx, hyperref, amsmath, url}
\usepackage[a4paper,text={16.5cm,25.2cm},centering]{geometry}
\usepackage{booktabs}
\usepackage{float}
\restylefloat{table}
\usepackage{authblk}

\author[1]{Rub\'en H. Garc\'ia-Ortega}
\author[2]{Juan J. Merelo Guerv\'os}
\author[3]{Pablo Garc\'ia S\'anchez}
\author[4]{Gad Pitaru}
\affil[1,4]{Badger Maps, San Francisco, CA, USA}
\affil[2]{Universidad de Granada and CITIC-UGR, Spain}
\affil[3]{Universidad de Cádiz and CITIC-UGR, Spain}

\hypersetup
{
  pdfauthor = {Rub\'en H. Garc\'ia-Ortega},
  pdftitle={Overview of \textit{PicTropes}, a film trope dataset},
  colorlinks=TRUE,
  linkcolor=black,
  citecolor=blue,
  urlcolor=blue
}

\title{Overview of \textit{PicTropes}, a film trope dataset.}
\date{23rd June 2018}

\begin{document}
\maketitle

\begin{abstract}
   From the database \textit{DBTropes.org}, we have created a dataset of films and the tropes that they use,
   which we have called \textit{PicTropes}.
   In this report we provide the descriptive analysis and a further discussion on this new dataset:
   The extracted features will help us decide the best values for a future recommendation system and content generator,
   whereas the analysis of the distribution functions that fit the best will help us interpret the relation
   between the films and the tropes that were found inside them.
   Additionally, we provide rankings of the  top-25 tropes and films, which will help us discuss and
   formulate questions to guide future extensions of the \textit{PicTropes} dataset.
\end{abstract}

\section{Introduction}

A trope can be defined as a recurring narrative device~\cite{baldick2015oxford};  it can be a
technique, a motif, an archetype or a \textit{cliché},
used by the authors to achieve specific effects that might vary from increasing the interest, surprising, recall familiarity,
entertaining, etc, in their creative works, such as books, films, comics or videogames.
Some tropes are broadly adopted, academically studied and promoted, such as
the \textit{Three-act Structure} formulated by Syd Field~\cite{field1982screenplay},
the \textit{Hero's Journey} studied by Vogler~\cite{vogler2007writer},
the \textit{McGuffin} popularized by Hitchcock~\cite{truffaut1985hitchcock} and
the \textit{Chekhov's Gun} developed by the Russian writer with the eponymous name~\cite{bitsilli1983chekhov},
but there are thousands of not-so-widely used tropes as well, discovered and catalogued everyday by professionals
and enthusiastic of the storytelling;
their study is organic, dynamic and extensive.
All of these tropes are described in a live wiki called \textit{TVTropes.org~}\cite{tvtropes}, that is being
collecting thousand of descriptions and examples of tropes from 2014 until now. As the data is fed by a community
of users, we could find the bias that popular films are better described and analysed in terms of the tropes than independent
films, and that popular tropes are more recognised than very specific ones.
The semantic network of knowledge behind \textit{TVTropes.org} is huge and complex; it massively links hierarchies of tropes
to their usage in creations for digital entertainment. The data, however, is only available through its web interface,
which is why, in order to make it usable by the scientific community, Kiesel~\cite{maltekiesel} extracted all
their data to a database so-called \textit{DBTropes.org}.\\

\textit{DBTropes.org} is released as a NTriples formatted RDF file that can be downloaded directly from their
official site~\cite{maltekiesel}.
The last release available was built in July 2016 and contains 2,1057,602 RDF statements,
a large amount of data that makes it hard to import it in a RDF visualization tool.

In this work, we extract part of its information
to a new dataset in JSON format, readable by most of the programming languages in a friendly manner,
that is called \textit{PicTropes} and contains just the films and the name of the tropes they use.\\

The goal of this report is to extract valuable statistical data from the dataset \textit{PicTropes}
that shall be used in further researches and experiments related to machine learning
and narrative generation. In particular, the results will be directly applicable to different researches
in the context of the PhD
\textit{Bio-inspired techniques for procedural generation of backstories in literature and open world videogames}.\\

Following the principles of the Open Science,
the dataset and the source files of the report are released under the
\textit{Attribution-ShareAlike 3.0 Unported} License (CC BY-SA 3.0)
in a public repository~\cite{ruben_tropes_open_data_2018}.
The values and graphs included in the current report are
dynamically calculated using pweave, scipy, numpy and matplotlib in order to allow the reproducibility.
The class that handles the Python code for the different calculations
can be found in the same repository.\\

In order to ease the data analysis we will use two data structures: a dictionary of films where the values are lists
of the tropes they use, analized in Section~\ref{sec:films-by-trope},
and the reverse dictionary, whose keys are the tropes and values are lists of films that
they are used in, analyzed in Section~\ref{sec:tropes-by-film}.
The information extracted from the dataset is discussed in Section~\ref{sec:discussion}
and Section~~\ref{sec:conclusions} summarizes the conclusions.

\section{Descriptive analysis of the number of tropes per film} \label{sec:films-by-trope}

\begin{figure}[H]
\center
\includegraphics[width= \linewidth]{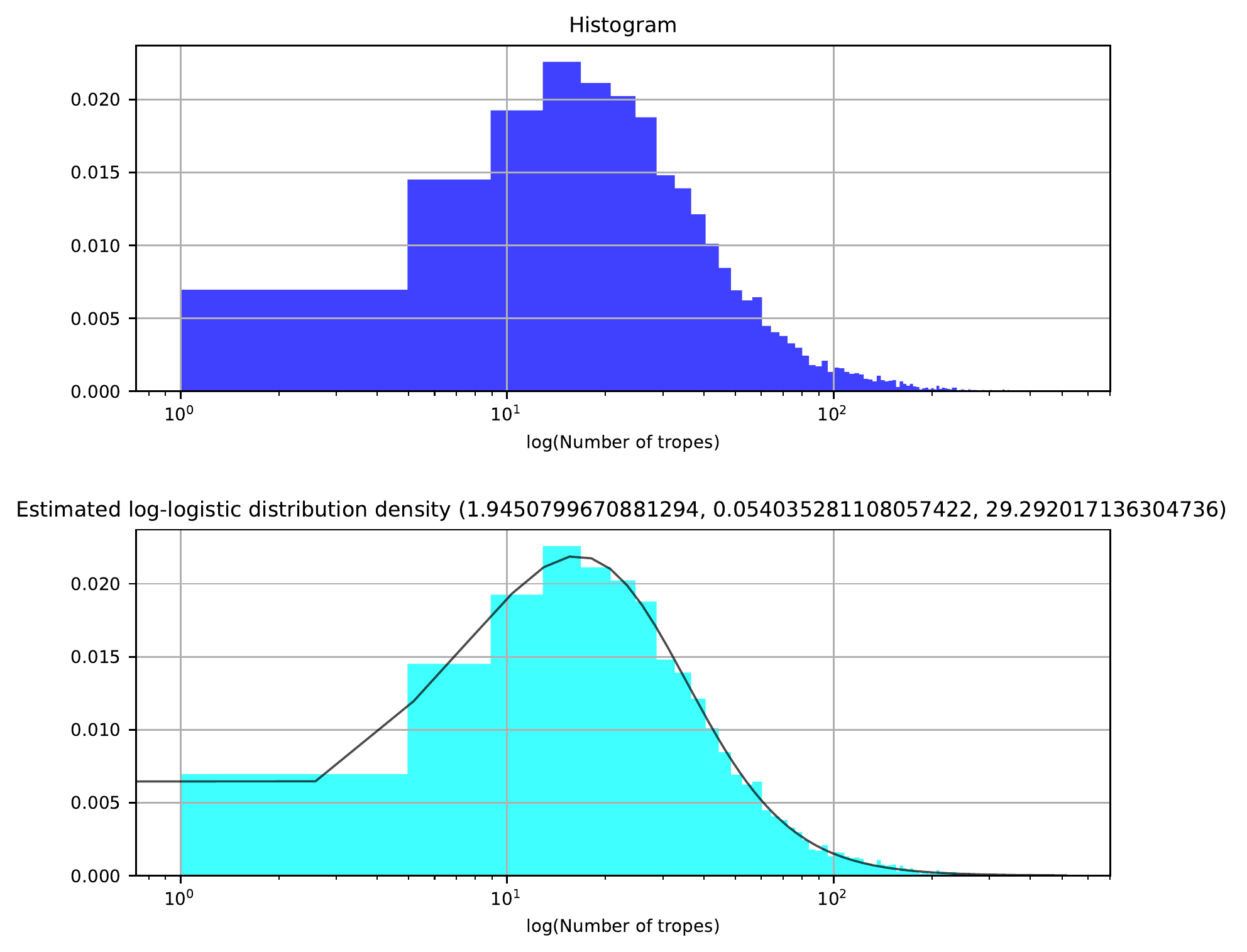}
\caption{Distribution of the number of tropes per film}
\label{fig:distribution:films}
\end{figure}

The dataset \textit{PicTropes} contains 5,925 films with the tropes that have been
found in each of them. The number of tropes in a film goes up to 515
(\textit{GuardiansOfTheGalaxy});
however, the great majority of films have just a few dozens of tropes
(43.433999999999997 on average and 29.0 on median).
As shown in Figure~\ref{fig:distribution:films} the data fits a
log-logistic distribution (\textit{location}=1.9450000000000001,
\textit{shape}=0.053999999999999999, \textit{scale}=29.292000000000002).
The features of the descriptive analysis can be found in Table~\ref{statistics-tropes-by-film}
whereas the top-25 ranking of films by the number of tropes is provided in Table~\ref{top25-films-table} for further analysis.

\begin{table}[H]
\centering
\caption{Descriptive analysis of the number of tropes per film}
\label{statistics-tropes-by-film}
\begin{tabular}{@{}lr@{}} \toprule
Number of films & 5,925 \\ \bottomrule \\
 & \textbf{Tropes per film} \\ \midrule
Minimum &  1 \\
Maximum &  515 \\
Mean &  43.434 \\
Median &  29.0 \\
Q1 &  16.0 \\
Q2 &  29.0 \\
Q3 &  52.0 \\
Variance & 2,133.35 \\
Skewness & 3.332 \\
Kurtosis & 17.373 \\ \bottomrule

\end{tabular}
\end{table}

\begin{table}[H]
\centering
\caption{Top-25 ranking of films by  the number of tropes}
\label{top25-films-table}
\begin{tabular}{@{}llr@{}}
\toprule
\textbf{Position} & \textbf{Film (short name)} & \textbf{N. tropes} \\ \midrule

1 & GuardiansOfTheGalaxy & 515 \\ 
2 & TheDarkKnightRises & 503 \\ 
3 & XMenDaysOfFuturePast & 478 \\ 
4 & CaptainAmericaTheFirstAvenger & 472 \\ 
5 & XMenFirstClass & 469 \\ 
6 & Thor & 427 \\ 
7 & SherlockHolmes & 411 \\ 
8 & TheLordOfTheRings & 398 \\ 
9 & PacificRim & 385 \\ 
10 & CaptainAmericaTheWinterSoldier & 382 \\ 
11 & WhoFramedRogerRabbit & 370 \\ 
12 & TheDarkKnight & 362 \\ 
13 & TronLegacy & 360 \\ 
14 & StarTrek & 353 \\ 
15 & StarTrekIntoDarkness & 347 \\ 
16 & Skyfall & 344 \\ 
17 & TheGodfather & 342 \\ 
18 & JurassicWorld & 332 \\ 
19 & Serenity & 331 \\ 
20 & BackToTheFuture & 330 \\ 
21 & Inception & 326 \\ 
22 & IronMan3 & 320 \\ 
23 & AustinPowers & 314 \\ 
24 & GalaxyQuest & 312 \\ 
25 & ThorTheDarkWorld & 302 \\ 

\bottomrule
\end{tabular}
\end{table}

\section{Descriptive analysis of the number of films per trope} \label{sec:tropes-by-film}

\begin{figure}[H]
\center
\includegraphics[width= \linewidth]{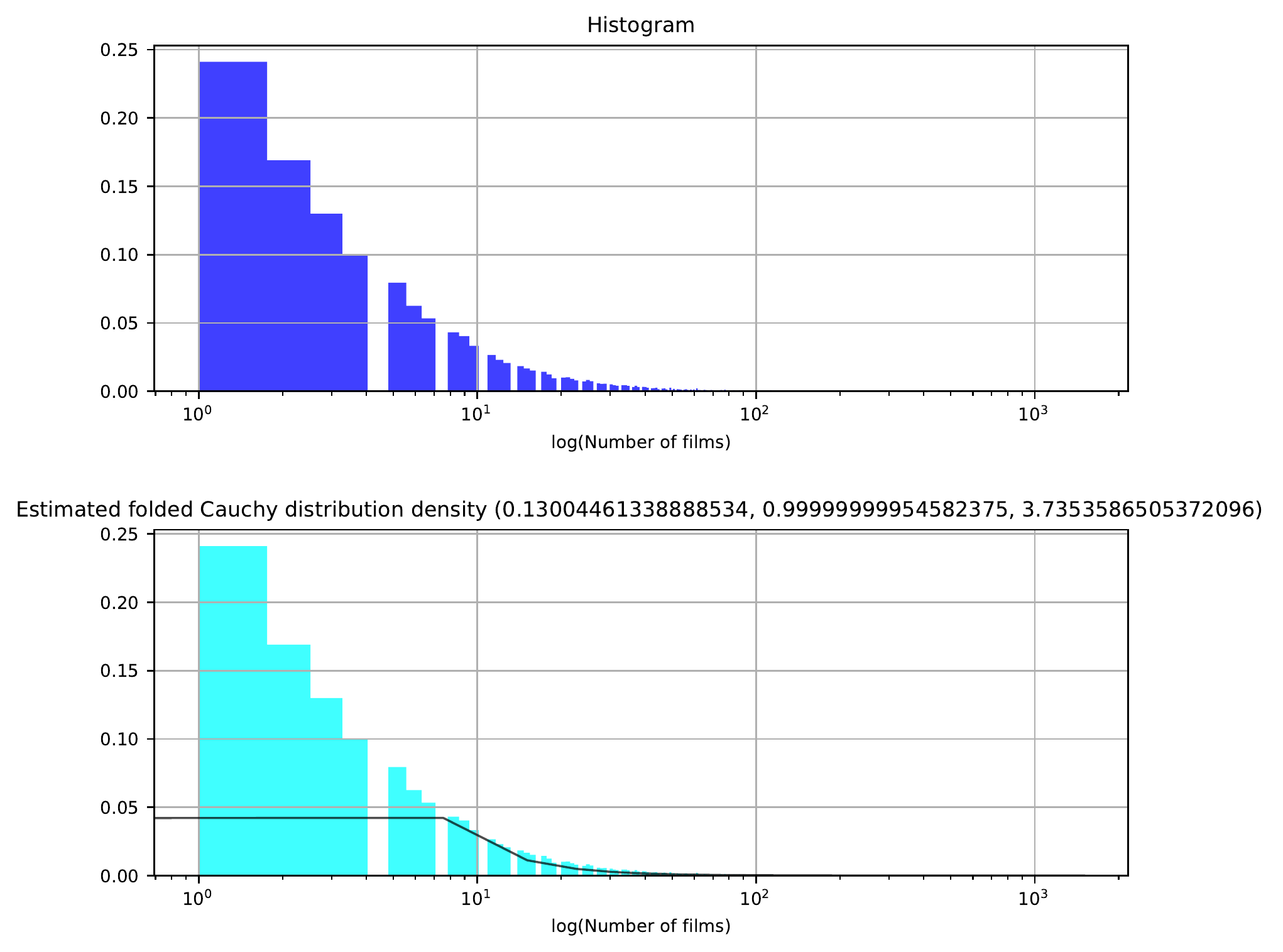}
\caption{Distribution of the number of films per trope}
\label{fig:distribution:tropes}
\end{figure}

The dataset \textit{PicTropes} contains 18,270 tropes with the films where they have been
found. The number of films where a specific trope appears goes up to 1502
(\textit{ShoutOut});
however, the great majority of tropes are only found in a few films
(14.086 on average and 5.0 on median).
As shown in Figure~\ref{fig:distribution:tropes} the data fits a
folded Cauchy distribution (\textit{location}=0.13,
\textit{shape}=1.0, \textit{scale}=3.7349999999999999).
The features of the descriptive analysis can be found in Table~\ref{statistics-films-by-trope}
whereas the top-25 ranking of tropes by the number of films they appear in is provided in
Table~\ref{top25-tropes-table} for further analysis.

\begin{table}[H]
\centering
\caption{Descriptive analysis of the number of films per trope}
\label{statistics-films-by-trope}
\begin{tabular}{@{}lr@{}} \toprule
Number of tropes & 18,270 \\ \bottomrule \\
 & \textbf{Films per trope} \\ \midrule
Minimum &  1 \\
Maximum &  1,502 \\
Mean &  14.086 \\
Median &  5.0 \\
Q1 &  2.0 \\
Q2 &  5.0 \\
Q3 &  12.0 \\
Variance & 1,464.794 \\
Skewness & 11.758 \\
Kurtosis & 245.951 \\ \bottomrule

\end{tabular}
\end{table}

\begin{table}[H]
\centering
\caption{Top-25 ranking of tropes by  the number of films}
\label{top25-tropes-table}
\begin{tabular}{@{}llr@{}}
\toprule
\textbf{Position} & \textbf{Trope (short name)} & \textbf{N. films} \\ \midrule

1 & ShoutOut & 1502 \\ 
2 & ChekhovsGun & 994 \\ 
3 & OhCrap & 823 \\ 
4 & DeadpanSnarker & 805 \\ 
5 & Jerkass & 784 \\ 
6 & Foreshadowing & 746 \\ 
7 & LargeHam & 724 \\ 
8 & BittersweetEnding & 697 \\ 
9 & TitleDrop & 634 \\ 
10 & BigBad & 612 \\ 
11 & MeaningfulName & 588 \\ 
12 & BerserkButton & 555 \\ 
13 & TheCameo & 542 \\ 
14 & WhatHappenedToTheMouse & 538 \\ 
15 & RunningGag & 524 \\ 
16 & TooDumbToLive & 521 \\ 
17 & FanService & 516 \\ 
18 & DownerEnding & 516 \\ 
19 & KarmaHoudini & 514 \\ 
20 & GroinAttack & 512 \\ 
21 & BrickJoke & 484 \\ 
22 & BookEnds & 473 \\ 
23 & MoodWhiplash & 460 \\ 
24 & KickTheDog & 455 \\ 
25 & PrecisionFStrike & 447 \\ 

\bottomrule
\end{tabular}
\end{table}

\section{Discussion} \label{sec:discussion}

The descriptive analysis of the data shows that, on average, a film contains 43.434
tropes and that a trope is contained in 14.086 films.
However, the distribution in both cases is long tailed.

In the case of the films, some of them include hundred of tropes, but the majority include just a few dozens.
That's clearly visible in the top-25 ranking, where the difference between
the top (\textit{GuardiansOfTheGalaxy})
and the bottom (\textit{ThorTheDarkWorld})
is 41.359~\%.
The reason why this difference is so big is not clear but, as mentioned in the introduction, it could be because
they actually use more tropes or because of the bias that the bigger the fandom is the more the analysis is performed and the more the
tropes are found.
It is specially
noticeable that 22 out of the 25 belong to the \textit{adventure} genre,
that 10 out of the 25 are about super-heroes of the comics,
that most of the films belong to this millennium and are super-productions.

In order to fully analyze the results we should study a possible correlation between the "attention" and the number of tropes.
External movie databases as \textit{IMDB} or \textit{Movielens} provide the popularity of the films, votes, genres and release dates;
with this extra information we could make solid statements that could help us to ignore tropes or find hidden ones.
Furthermore, knowing the rating of the films would help apply Machine Learning techniques to predict the intrinsic quality
of a set of tropes.

When we analyze the tropes, the effect of the long tail is even more noticeable and the difference between
the top (\textit{ShoutOut})
and the bottom (\textit{PrecisionFStrike})
in the top-25 ranking is
70.24~\%.
The trope \textit{Shout out}, deliberate allusions to other sources of
inspiration, is by far the most used and appears in the
25.35~\% of the analyzed films.
Looking at the long-tailed distribution function and the big differences in the top-25,
we could ask ourselves new interesting questions:
is the variance so considerable because of the popularity in the films or in the fandom?
Is the top-25 so widely discovered in the films because they are the \textit{master} formula of the entertainment
or it is because they are easy to apply in every story? In order to answer these questions, again, we would
require extra information, and, perhaps, a good ontology of tropes by different features.

Finally, it is important to remark that the last dump of \textit{DBTropes.org} was extracted in 2016. With newer data
the analysis could've been updated as we could've consider newer films and tropes.

\section{Conclusions} \label{sec:conclusions}

The goal of the following report is to provide the descriptive analysis and a further discussion on the dataset
\textit{PicTropes},
that links 5,925 films
with 18,270 tropes from the database \textit{DBTropes.org}.

In order to achieve the goal, two analysis have been performed, one on the tropes by film
and other on the films by trope. Both analysis included statistical descriptive analysis,
probability distribution fitting and a list of the top elements.

The results show the number of tropes in a film goes up to 515
(\textit{GuardiansOfTheGalaxy});
however, the average of tropes by film is (43.433999999999997) and the data fits a
log-logistic distribution (\textit{location}=1.9450000000000001,
\textit{shape}=0.053999999999999999, \textit{scale}=29.292000000000002).
This distribution is characterized by a long tail and by a mode that is close to the minimum, that means that
all the films tend to have just a dozen of tropes, and a great minority have hundred of them.
In the discussion, we point out the fact that most of the films with
more tropes are super-productions, a dozen of years old at maximum and in the adventure genre.

Regarding the tropes, the number of films by trope goes up to 1502
(\textit{ShoutOut});
however, the average of films by trope is (14.086) and the data fits a
folded Cauchy distribution (\textit{location}=0.13,
\textit{shape}=1.0, \textit{scale}=3.7349999999999999).
This distribution is also characterized by a long tail and by a mode that is close to the minimum, that means that
all the tropes tend to be discovered in a few films, but a great minority are present un a huge number of films.
In the discussion, we show the need to have a good ontology of tropes so we can determine the root of the popularity
of a small minority of tropes: if they are just easy to use or if their use correspond to other creative decisions.

This analysis is useful because it can be used together with the dataset \textit{PicTropes} to
automatically generate plots that follow the canonical appearance of tropes based on empirical observations,
by using data mining and machine learning techniques. The results will be directly applicable to different researches
in the context of the PhD
\textit{Bio-inspired techniques for procedural generation of backstories in literature and open world videogames}.\\

Finally, as future work, we could avoid some current limitations of the dataset \textit{PicTropes} by adding films meta information
(as genre ontology, votes, popularity or release date) and tropes meta information (as the ease of application
and the effects of this application), so the automatic generation could lead to plots with higher quality.
It would imply retrieving data from movie databases using different techniques
and prepare it so it matches with the tropes and films uniquely.
Regarding \textit{DBTropes.org}, as its last dump was extracted in 2016, it could be interesting to contribute
to the project in order to have a newer version, with the latests additions and contributions.

\bibliographystyle{ieeetr}
\bibliography{report}

\end{document}